\documentstyle [prl,aps,epsf,multicol]{revtex}

\begin{document}
\newcommand{\fig}[2]{\epsfxsize=#1\epsfbox{#2}}
\title{Photocurrent in conjugated polymers \\}
\author{Baruch Horovitz and Ehoud Pazy}
\address{Department of Physics and Ilze Katz center for
nanotechnology, Ben-Gurion University of the Negev, Beer-Sheva
84105, Israel}

\maketitle

\begin{abstract}
Nonlinear photocurrent carriers in conjugated polymers, such as
polarons, bipolarons and solitons, are considered at low photon
energies where a tunnelling process is necessary. We show that
polarons usually dominate the photocurrent $I$ due to a novel
electric field assisted tunnelling for which $\ln I\sim
-E^{-2/3}$. For near degenerate polymers an electric field $E$
which exceeds the confinement potential and frequencies above
twice the soliton energy, soliton tunnelling is favored.
Photocurrent data can then be used to identify the remarkable
phenomenon of soliton conduction.

\end{abstract}


\begin{multicols}{2}

Photoexcitation of charge carriers in conjugated polymers is of
considerable interest for determining types of charge carriers and
for probing novel conduction mechanisms. Considerable theoretical
attention has been given to polyacetylene (PA)
\cite{su,ball,sethna,heeger} which has degenerate ground states
(i.e. by inverting the dimerization sign) predicting photocurrent
carried by charged solitons. Experimentally, however, significant
photocurrent has been observed in non-degenerate conjugated
polymers such as polydiacetylene (PDA) \cite{siddiqui,moller}.

In a non-degenerate polymer the dimerization pattern in the ground
state is unique, being fixed by the polymer structure. The charge
carriers are expected to be \cite{heeger} either polarons or
doubly charged bipolarons; the continuum model for polarons and
bipolarons has been considered in detail
\cite{brazovskii,campbell,onodera}. Reversing the sign of the
dimerization leads to a local minimum (at least for near
degenerate systems) which is higher by an energy of $\alpha$ per
unit length. A soliton is an excitation which interpolates between
states with opposite signs of dimerization; hence generating a
soliton pair leads to an excess energy linear in the separation
$R$ of the solitons, i.e. a confinement potential $\alpha R$. In
contrast, polarons or bipolarons are local deformations of the
ground state, hence a polaron pair has no confinement potential.
We note that interchain coupling acts as a confinement potential,
hence even PA with degenerate minima should be considered as a
near-degenerate polymer.

Some properties of PDA serve us as a prime example for
illustrating phenomena in a non-degenerate polymer. PDA has chains
of the form $[=RC-C \equiv C-CR=]_n$ where R is one of a number of
side groups with which the monomer can be synthesized. Band
structure calculations \cite{bredas} show that the states near the
gap are essentially $\pi _z$ orbitals. If these orbitals were
removed the underlying structure would be $[-RC-C=C-CR-]_n$, i.e.
3 bonds per carbon. If $a$ is the mean spacing of carbons, the
electron spectrum in an extended zone scheme has gaps at $\pm
n\pi/4a$ with $n=1,2,3$ since the unit cell has 4 carbons. Filling
in one electron per Carbon for the $\pi _z$ orbitals up to
wavevectors $\pm \pi/2a$ leads therefore to a gap at the Fermi
level, a so called "extrinsic gap" $\Delta_e$. In addition, the
$\pi _z$ orbitals tend to dimerize, i.e. their overlap between
nearest carbons is alternating. This dimerization increases the
gap at the Fermi level so that the gain in electron energies
overcomes the cost in the lattice distortion. For PDA, the usual
acetylenic ground state is then formed with the above triple bond.
It is possible, however, depending on the side groups R, on
temperature or on external strain, that the ground state will
favor the opposite sign for this dimerization, leading to the
butatrienic form $[-RC=C=C=CR-]_n$. In fact, some data supports a
PDA butatrienic form produced in a different photopolymerization
procedure, which transforms into an acetylenic form when annealed
above $\approx 350K$ \cite{kuriyama}. Solitons in PDA interpolate
between these states, i.e. $[...-RC=C=C=CR-R{\dot C}-C \equiv
C—CR=...]$. The central carbon ${\dot C}$ has only three bonds,
i.e. it acquires a localized orbital with intragap energy. If a
phase transition in PDA is indeed posible, then at the critical
temperature $\alpha =0$ and solitons can lead to photocurrent
\cite{sethna2}. As we show below, even a small $\alpha$ can
demonstrate the remarkable phenomenon of soliton conductance.

Experimental data on PDA \cite{siddiqui} with currents measured
down to $10^{-16}A$ shows a steep photocurrent edge at photon
energy of 0.8eV which is well below either the band gap of 2.4eV,
or the exciton level at 2eV. Further data on PDA with a different
side group \cite{moller} has shown that thermal annealing allows
for measurable photocurrents at 2.2eV, with an exciton level at
2.3eV; the current sensitivity in this case was about $10^{-13}A$.

In the present work we study photocurrent due to tunnelling into a
variety of charge carriers -- polarons, bipolarons and solitons.
We show that at low photon energies $\hbar \omega <2E_p$ polarons
usually dominate, where $E_p$ is the polaron formation energy. The
electric field allows for a novel tunnelling route where weak
lattice deformations are formed at large separation $R_p$ such
that upon charge transfer shallow polarons are formed; the energy
gain $eER_p$ compensates for the missing formation energy
$2E_p-\hbar \omega$. This process is likely to be effective also
in semiconductors with higher dimensionality. We also show that
bipolarons can be directly photoproduced, though with a smaller
probability. Finally we show that the external electric field $E$
can overcome the confinement potential when $eE>\alpha$ and allows
for charged soliton creation. We find that for $2E_s<\hbar \omega
< 2E_p$, and above the threshold field $\alpha/e$ soliton
formation is favored, where $2E_s$ is the soliton pair energy (in
addition to the $\alpha R$ term). This process can serve as a
clear identification of soliton conduction, adding to intriguing
PA data \cite{heeger}.

We consider first photocurrent due to polarons at $\hbar \omega
<2E_p$. The tunnelling process is parameterized by two local
deformations of the dimerization pattern separated by distance
$R$, each deformation yields a bound state of size $1/\kappa$.
These localized states are pulled from the valence and conduction
band edges at $\pm \Delta_0$, respectively. We assume that direct
electron-electron (e-e) interactions are embedded in the
parameters such as $\Delta_0$, $E_p$ and $\kappa$; where needed we
address e-e interactions explicitly. For the purpose of
illustration, we show the electron level structure in Fig. 1 for
the electron-lattice (e-l) model in the absence of e-e
interactions.

The initial state potential $V_0(\kappa,R)$ describes the deformed
lattice with the electron occupation following adiabatically that
of the ground state. The excited state corresponds to a dipole
allowed transition of one electron into a state with energy higher
by $2\omega_0$, e.g. for $\kappa =0$ we have
$2\omega_0=2\Delta_0$. The excited state can relax adiabatically
into a two polaron state $P^+P^-$ with energy minimized at some
$\kappa =\kappa_p$, i.e. the polaron energy $E_p(\kappa)$ relaxes
to $E_p\equiv E_p(\kappa_p)$. In the e-l model the initial state
of each deformation has the lower level (Fig. 1a) doubly occupied
while the upper level is empty; the excited state has one more
electron in the upper state ($P^-$) or one less in the lower state
($P^+$).

\begin{figure}[htb]
\centerline{ \fig{8cm}{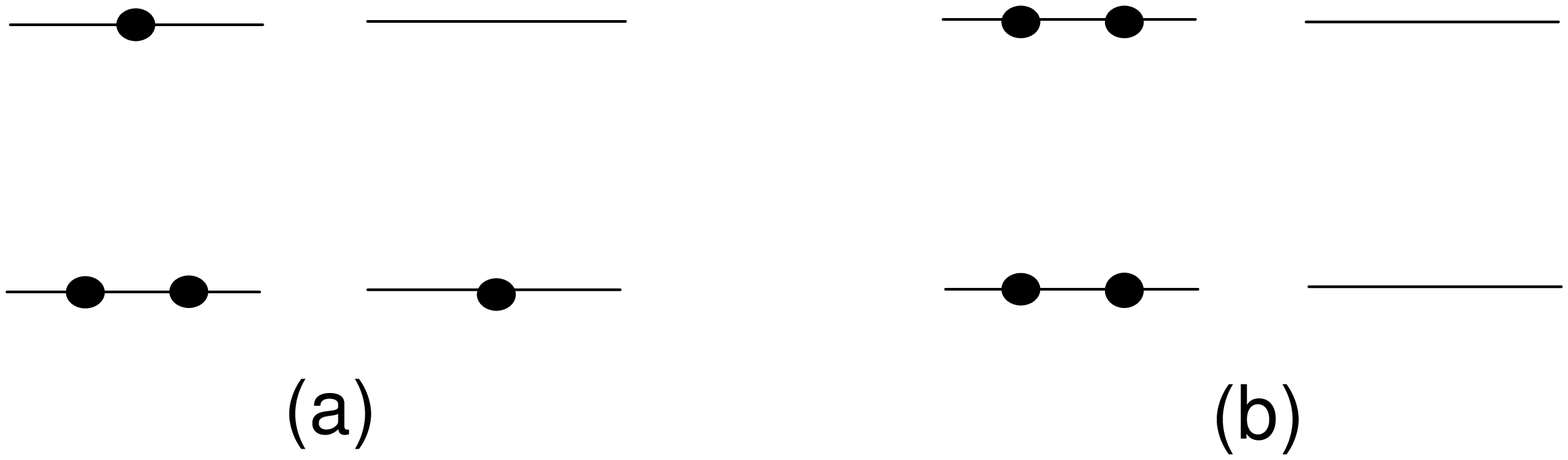} }
\caption{Two localized states with intragap energies $\pm
\omega_0$ (continuum states starting at $\pm \Delta_0$,
$\Delta_0>\omega_0$, are not shown). The dots represent occupation
of these levels in the e-l model for (a) a separated polaron pair
$P^-P^+$, (b) a separated bipolaron pair $B^+B^-$.}
\end{figure}

The condition for having a localized state with well defined
charge is (i) $R\gg 1/\kappa$ and (ii) that $E$ does not mix the
state with continuum states, i.e. the energy shift of the
localized state $\approx eE/\kappa$ is small relative to the
excitation energy into the continuum, i.e. $eE/\kappa \ll
\Delta_0-\omega_0$. We claim that for small $\kappa$ the latter is
$\Delta_0-\omega_0\approx \Delta_0\kappa^2\xi^2$ where
$\xi=v_F/\Delta_0$ and $v_F$ is the Fermi velocity in the absence
of an energy gap. For the noninteracting case this is seen by
continuing analytically the momentum $k\rightarrow i\kappa$ in the
spectrum $\omega_0=\sqrt{\Delta_0^2-v_F^2\kappa^2}\approx
\Delta_0(1-\frac{1}{2}\kappa^2\xi^2)$. In the final paragraph
before the conclusions we show that to first order in Coulomb
interactions the $\kappa^2$ form is maintained. This is rather
surprising since the Hartree term by itself gives a $\sim \kappa$
term \cite{heeger}. The perturbation parameter being
$e^2\kappa/\Delta_0$ implies that higher order terms lead to
higher powers of $\kappa$.

The condition $eE/\kappa \ll \Delta_0-\omega_0$  becomes then
$\kappa\gg \kappa_m$ where
\begin{equation} \label{km}
\kappa_m=(eE\xi/\Delta_0)^{1/3}/\xi \,.
\end{equation}
Since typically $\Delta_0\approx 1eV$, $\xi\approx 1nm$ and
$E\lesssim 10^5V/cm$ we have $eE\xi/\Delta_0\lesssim 10^{-2}$
which serves as a small parameter. This condition for localized
charges is extremely important -- it allows the excited state
potential to gain an electric field energy
$V_{ex}(\kappa,R)=2E_p(\kappa)-eER$, hence the energy gain $eER$
facilitates tunnelling even if $\hbar \omega <2E_p$. The initial
state potential has neutral components, hence
$V_0(\kappa,R)=2E_p(\kappa)-2\omega_0(\kappa)$.

Perturbation theory in the exciting photons $\sim e^{i\omega t}$
\cite{sethna,horovitz} in the adiabatic limit ($\hbar \omega,
2\Delta_0$ large compared with a typical phonon energy
$\hbar\omega_{ph}$) shows that the electron transition occurs at
$\kappa, R$ such that $V_0(\kappa, R)=V_{ex}(\kappa, R)-\hbar
\omega$. The tunnelling barrier changes then from $V_0(\kappa, R)$
to $V_{ex}(\kappa, R)-\hbar \omega$ at the crossing point. The
dynamics is dominated by the ion kinetic energy; the space
dependent  dimerization pattern $\Delta(x;k,R)=\Delta_p(\kappa,
x-R/2)+\Delta_p(\kappa, x+R/2)$, where $\Delta_p(\kappa, x)$ is
the single polaron pattern, leads to the kinetic energy
\begin{equation}\label{Ek}
\int dx \frac{{\dot \Delta}^2}{\pi \lambda v_F\omega_{ph}^2}=
M_1(\kappa){\dot R}^2/2 +M_2(\kappa){\dot \kappa}^2/(2\kappa^4)
\end{equation}
where dot is a time derivative and $\lambda$ is the
electron-phonon coupling. $M_1(\kappa)$ is the mass for polaron
motion while $M_2(\kappa)$ is the mass for the polaron size
$1/\kappa$ modulations. Note that $\Delta_p(\kappa, x)$ is
symmetric in $x$ hence $\partial_R$ terms are antisymmetric and
the coefficient of a cross term ${\dot R}{\dot \kappa}$ vanishes.

Consider first a tunnelling trajectory such that initially
$\kappa$ increases (at small $R<\xi$) until crossing into the
excited potential occurs. The matrix element for the transition
involves the photon field and a wavefunction overlap which is
large for $R<\xi$. The tunnelling continues then on the excited
state with increasing $R$, reducing the excited state energy with
the $-eER$ term until it vanishes at $R_1=(2E_p-\hbar \omega)/eE$
(Fig. 2a). The condition $R_1\gg \xi$ is satisfied if $2E_p-\hbar
\omega$ is not too small, e.g. $\gtrsim 0.1eV$ and with
$\xi\approx 1nm$ one needs $E \ll 10^6 V/cm$. Most of the
trajectory is then in the regime $R\gg \xi$ justifying the use of
the form $V_{ex}(\kappa,R)=2E_p-eER$ and the use of effective
masses. The tunnelling rate along the $R$ trajectory under the
barrier $2E_p-eER-\hbar \omega$ (curve p in Fig. 2a) is given by a
WKB form $\exp\{-2\int_0^{R_p}[M_p(2E_p-eER-\hbar
\omega)]^{1/2}dR\}$ where $M_p=M_1(\kappa_p)$ is the polaron mass;
hence
\begin{eqnarray}\label{Gamma1}
\Gamma_1\sim e^{ -\frac{4\sqrt{M_p}}{3eE}(2E_p-\hbar
\omega)^{3/2}} \,.
\end{eqnarray}

This exponent involves two large parameters; first an electric
field term, $\Delta_0/eE\xi > 10^2$, and second an adiabatic
parameter $\sqrt{M_pv_F^2/\Delta_0}\sim
\Delta_0/\omega_{ph}\approx 10$. Hence $\Gamma_1$ is extremely
small.

\begin{figure}[htb]
\centerline{ \fig{8cm}{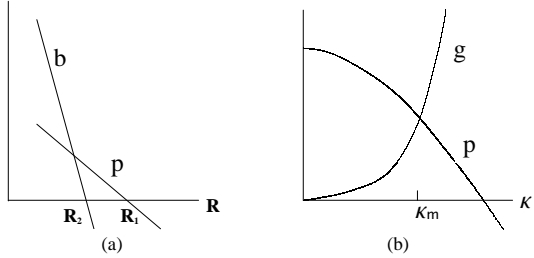} }
\caption{Energy barriers for (a) tunnelling along $R$ for polarons
(curve $p$ with $2E_p-\hbar\omega-eER$) and for bipolarons (curve
$b$ with $2E_b-\hbar\omega-2eER$), and (b) tunnelling along
$\kappa$: the initial deformed ground state energy (curve $g$ with
$2E_p(\kappa)-2\omega_0(\kappa)$) crosses into the polaron state
(curve $p$ with $2E_p(\kappa )-eER-\hbar\omega$) at the minimal
$\kappa$ value of $\kappa_m$.}
\end{figure}

To find the optimal tunnelling trajectory in the $R,\kappa$ plane
and avoid the strong suppression of tunnelling along $R$ (Eq.
\ref{Gamma1}) we consider the opposite extreme where $\kappa$
increases from $0$ at a large fixed $R$. This corresponds to
nucleating two lattice deformations at a large distance $R$ with
initial energy $2E_p(\kappa)-2\omega_0(\kappa)$ which crosses the
excited state energy at some $\kappa$ (Fig. 2b). The price to pay
in this process is that the matrix elements involve the overlap of
localized wavefunctions separated by a large $R$, i.e. $\ln \Gamma
\approx -2\kappa R$. As discussed above, the minimal $\kappa$ that
is consistent with localized charges is $\kappa_m$, hence we
choose to nucleate the deformations at a distance $R_p$ such that
the crossing in Fig. (2b) is at $\kappa_m$, i.e.
\begin{equation}\label{Rp}
2E_p(\kappa_m)-2\omega_0(\kappa_m)=2E_p(\kappa_m)-\hbar
\omega-eER_p \,.
\end{equation}
Since $\kappa_m$ is small $R_p\approx (2\Delta_0-\hbar
\omega)/eE$, hence the tunnelling rate is
\begin{equation}\label{Gammap}
\Gamma_p \sim e^{-4(1-\frac{\hbar
\omega}{2\Delta_0})(\frac{\Delta_0}{eE\xi})^{2/3}}
\end{equation}
$\Gamma_p$ involves also the tunnelling across small $\kappa$
values of Fig. 2b which contributes a small correction within the
exponent in Eq. (\ref{Gammap}). We note that if Coulomb energies
would give an excitation energy $\sim \kappa$ (e.g. in
semiconductors with higher dimensionality) then $\kappa_m\sim
E^{1/2}$ and $\ln I\sim (\Delta_0/eE\xi)^{1/2}$. We also note that
choosing $R<R_p$ for Fig. 2 implies a higher $\kappa$ so that the
above $\kappa_mR_p$ is the minimal value.

The result for $\Gamma_p$ is remarkable -- it reduces the large
$\Delta_0/eE\xi$ factor in Eq. (\ref{Gamma1}) to its $2/3$ power
and also avoids the large adiabatic parameter. Furthermore, this
process is much more efficient than the Franz-Keldysh process
\cite{FK} which is the photon assisted $\kappa=0$ process, i.e.
tunnelling between extended states of the valence and conduction
bands with $\ln \Gamma \sim -
(1-\hbar\omega/2\Delta_{0})^{3/2}\Delta_0/eE\xi$. We conclude that
the process leading to Eq. (\ref{Gammap}) is the most efficient
one for generating polarons. This is a most remarkable process --
the system prepares localized states at a large separation
(typically $R_p>100nm$) so as to accommodate an eventual transfer
of charge.

We consider next tunnelling into bipolarons. The electric field
breaks inversion symmetry and allows charge transfer between
polarons, leading to doubly charged bipolarons. The excitaion
potential is $V_b(\kappa,R)=2E_b(\kappa)-2eER$ where $2E_b$ is the
creation energy of two bipolarons; in the e-l model
$2E_b(\kappa)=2E_p(\kappa)+2\omega_0$ (Fig. 1b). If tunnelling
along R had dominated, then direct tunnelling into bipolarons
would have dominated at low frequencies. To see this, note that
$V_b(\kappa, R)-\hbar\omega$ crosses $0$ at $R_2=(2E_b-\hbar
\omega)/2$ ($E_b$ here is the minimized $E_b(\kappa)$) which is
smaller than $R_1$ (Fig. 2a) if $\hbar \omega<4E_b-2E_p$, or
$\hbar \omega\lesssim 0.4\Delta_0$ in the e-l model. Tunnelling
directly into bipolarons eliminates the more costly tunnelling
range between $R_2$ and $R_1$ at the expense of an $e^{-2\kappa
R_2}$ factor for the second charge transfer. Hence, bipolarons
would dominate at low $\hbar \omega$.

Within the much more efficient process of tunnelling along
$\kappa$ at a large $R$ bipolarons are more efficiently generated
when $R=R_b$ is chosen such that the bipolaron curve
$2E_b(\kappa)-2eER-\hbar \omega$ crosses the initial energy
$2E_p(\kappa)-2\omega_0(\kappa)$ at the minimal $\kappa_m$, hence
$R_b\approx (2\Delta_0-\frac{1}{2}\hbar \omega)/eE$. The polaron
excitation energy is then below the bipolaron one, but since
$-eER$ is inefficient at $\kappa<\kappa_m$ the polaron curve would
also cross near $\kappa_m$. The transfer of two charges involves
now a $e^{-4\kappa R}$ factor, hence $\Gamma_b \ll \Gamma_p$ with
\begin{equation}\label{Gammab}
\Gamma_b \sim e^{ -8(1-\frac{\hbar \omega}{4\Delta_0})
(\frac{\Delta_0}{eE\xi})^{2/3}}\,.
\end{equation}

We consider next photocurrent of charged solitons which has been
studied in detail for the degenerate PA case
\cite{su,ball,sethna,heeger}. In the non-degenerate case, or for
PA with interchain coupling, the electric field must exceed a
threshold value to overcome the confinement potential. Solitons
being topological objects must be generated by $R$ tunnelling --
there is no $\kappa$ parameter or a local deformation which can
generate a single soliton from the ground state. The excitation
energy of a soliton pair $S^+S^-$ is then $V_s(R)=2E_s+\alpha
R-eER$ allowing tunnelling only for $eE>\alpha$. The tunnelling
terminates at $R_s=(2E_s-\hbar\omega)/(eE-\alpha)$ which is
assumed to be large, $R_s\gg \xi$. The tunnelling rate into
solitons with mass $M_s$, for $\hbar \omega <2E_s, eE>\alpha$, is
then
\begin{equation}\label{Gammas}
\Gamma_s\sim \exp [-\frac{\sqrt{8M_s}}{3(eE-\alpha)}(2E_s-\hbar
\omega)^{3/2} ]\,.
\end{equation}
The formation of $S^+S^-$ leaves a metastable state which can then
spontaneously form a pair of solitons without a photon. $\Gamma_s$
is extremely small (comparable to Eq. \ref{Gamma1}) unless
$\hbar\omega$ is close to $2E_s$. Soliton photocurrent may then
dominate the polaron one only if the range $E_s<\hbar\omega<E_p$
exists. In the e-l model with weak $\alpha$ one has $E_s\approx
0.6\Delta_0<E_p\approx 0.9\Delta_0$ ($E_s$ is defined so that a
soliton pair energy separated by $R\gg\xi$ is $2E_s+\alpha R$).

 We consider therefore the range
$2E_s<\hbar\omega<2E_p$. The polaron photocurrent is still given
by Eq. (\ref{Gammap}) while soliton tunnelling can occur at
$R<\xi$ where $\alpha$ and  $E$ are ineffective. The result is
then just as in the $\alpha=0$ system \cite{sethna} $\ln \Gamma
\approx -\Delta_0/\omega_0$, with weak $E$ dependence. We conclude
that soliton photocurrent dominates over that of polarons when
$2E_s<\hbar\omega<2E_p$ {\em and} $E>\alpha/e$. This is achievable
if $\alpha/e\lesssim 10^6V/cm$ or $\alpha=4\Delta_e\Delta_0/\pi
\lambda v_F$ \cite{onodera} yields $\gamma\equiv \Delta_e/\lambda
\Delta_0\lesssim 0.1$. Hence varying $E$ changes Eq.
(\ref{Gammap}) at $E>\alpha/e$ into a larger and weakly $E$
dependent tunneling rate, exhibiting a sensitive tool for
identifying the intriguing soliton conduction.

Finally, we present a somewhat technical section, evaluating the
effect of Coulomb interactions to first order on the excitation
energy of a polaron, e.g. $P^+$, leading to the important criteria
for $\kappa_m$ in Eq. (\ref{km}). The e-l system with
(non-uniform) electron transfer between nearest neighbors has
particle-hole symmetry, i.e. each eigenstate $\phi(n)$ at site $n$
with energy $E$ allows an eigenstate $(-)^n\phi(n)$ with energy
$-E$. Completeness relation for the polaron $P^+$ state (right
side of Fig. 1a) yields for the charge density of each spin
$\rho_{\uparrow}(n)=\frac{1}{2}$,
$\rho_{\downarrow}(n)=\frac{1}{2}-\phi_b^2(n)$ where $\phi_b(n)$
is the lower energy localized eigenstate (Fig. 1). A Hubbard
interaction, to first order is
$\sum_n(\rho_{\uparrow}(n)-\frac{1}{2})(\rho_{\downarrow}(n)-\frac{1}{2})=0$;
similarly all interactions with range of even sites vanish.

Consider next nearest neighbor interaction ${\cal H}_V$ with
coupling $V$. Its average yields direct and exchange terms,
\begin{eqnarray}\label{Hv}
\langle {\cal H}_V\rangle =&&\case{1}{2}V\sum_n\{(\rho(n)-
\case{1}{2})(\rho(n+1)-\case{1}{2})\nonumber\\
&& - |S+\phi_b^*(n)\phi_b(n+1)|^2-|S|^2\}
\end{eqnarray}
where $\rho(n)=\rho_{\uparrow}(n)+\rho_{\downarrow}(n)$ and
$S=\sum_{\alpha}\phi_{\alpha}^*(n) \phi_{\alpha}(n+1)$ sums on all
occupied continuum states (per spin) with energy up to
$-\Delta_0$. The 2nd term above is the exchange for $\uparrow$
spin while the 3rd is for $\downarrow$ spin. The summation can be
done in a continuum limit \cite{onodera} where $x=na$ and
$f_{\alpha},g_{\alpha}$ are wavefunctions on the even and odd
sites, respectively. For small $\kappa$ ($1/\kappa$ is the range
of the localized state) we obtain from \cite{onodera}
\begin{equation}\label{suma}
S=\frac{\Delta_e-\Delta_0}{2\pi \lambda v_F}-(1-\case{1}{2}\gamma
\xi \kappa)f_b(x)g_b(x)+O(\kappa^2)
\end{equation}
and the contribution to the polaron energy becomes
\begin{equation}\label{Hvp}
\langle {\cal H}_V\rangle_p =V_1 +Va\frac{\Delta_e-\Delta_0}{2\pi
\lambda v_F}(1-\frac{4}{\pi}\gamma \xi \kappa)+O(\kappa^2)
\end{equation}
where $V_1/N=-Va^2[(\Delta_e-\Delta_0)/2\pi \lambda v_F]^2$ is the
correction to the ground state energy per site. Note that the
direct term $\frac{1}{2}V\sum_nf_b^2(n)g_b^2(n)\sim O(\kappa)$
cancels with with the exchange term [Ref. \cite{heeger} presents
just the direct term which, as shown here, is insufficient].

The polaron $P^+$ excited state involves an electron transfer from
the top of the valence band with an extended wavefunction
$\phi_v(n)$ into the lower localized state (Fig. 1a) which becomes
doubly occupied. The direct term becomes
$\frac{1}{2}V\sum_n\phi_v^2(n)\phi_v^2(n+1)\rightarrow 0$ for an
extended system, the $\uparrow$ exchange is the same as in Eq.
(\ref{Hv}) while the $\downarrow$ exchange becomes
$|S+\phi_b^*(n)\phi_b(n+1)-\phi_v^*(n)\phi_v(n+1)|^2$. Using
\cite{onodera} $\int dx f_v^*(x)g_v(x)=\frac{1}{2}+O(\kappa^2)$ we
obtain that the excited polaron energy is the same as that of the
polaron to order $\kappa^2$. We have also checked this conclusion
within the continuum model with e-e interactions. Hence the
excitation energy is $O(\kappa^2)$, the same as in the e-l model.

In conclusion, we have shown that polarons usually dominate the
photocurrent when $\hbar\omega<2E_p$ with the tunnelling rate from
Eq. (\ref{Gammap}), i.e. $\ln I \sim -E^{-2/3}$. However, for
$2E_s<\hbar\omega<2E_p$ {\em and} $E>\alpha/e$ solitons dominate
the photocurrent. This is achievable for near-degenerate polymers,
i.e. a small $\alpha$. Furthermore, the intriguing possibility
that PDA has a phase transition at which the confinement parameter
$\alpha$ vanishes \cite{kuriyama} can be sensitively tested by the
field dependence of photocurrent data.

Acknowledgements: We thank Doron Cohen and Amir Berman for stimulating
discussions. This research was supported by THE ISRAEL SCIENCE FOUNDATION
founded by the Israel Academy of Sciences and Humanities and by a
German-Israeli DIP project.

\end{multicols}

\end{document}